\documentclass[twocolumn]{aastex62}
\usepackage[T1]{fontenc}
\usepackage{color}
\usepackage{amsmath,amstext}
\usepackage{hyperref}
\usepackage{natbib}

\newcommand\beq{\begin{equation}}
\newcommand\eeq{\end{equation}}

\begin{document}

\title{DISKS OF STARS IN THE GALACTIC CENTER TRIGGERED BY  TIDAL DISRUPTION EVENTS }

\author[0000-0002-3635-5677]{Rosalba Perna} 
\affil{Department of Physics and Astronomy, Stony Brook
  University, Stony Brook, NY 11794-3800, USA}
\affil{Center for Computational Astrophysics, Flatiron Institute, New York, NY 10010, USA} 

\author[0000-0001-7113-723X]{Evgeni Grishin}
\affil{School of Physics and Astronomy, Monash University, Clayton, VIC 3800, Australia}
\affil{OzGrav: Australian Research Council Centre of Excellence for Gravitational Wave Discovery, Clayton, VIC 3800, Australia}

\begin{abstract}

In addition to a supermassive black hole (SMBH),  the central parsec of the Milky Way hosts over a hundred of massive, high velocity young stars whose existence, and organisation of a subset of them in one, or possibly two, mis-aligned disks, is puzzling. Due to a combination of low medium density and strong tidal forces in the vicinity of Sgr A*, stars are not expected to form. Here we propose a novel scenario for their in-situ formation:
a jetted tidal disruption event (TDE) from an older wandering star triggers an episode of positive feedback of star formation in the plane perpendicular to the jet, as demonstrated via numerical simulations in the context of jet-induced feedback in galactic outflows. An over-pressured cocoon surrounding the jet shock-compresses clumps to densities high enough to resist the SMBH tidal field.
The TDE rate of $10^{-5}-10^{-4}$~yr$^{-1}$ per galaxy, out of which a few percent events are jetted, implies a jetted TDE event per galaxy to occur every few million years. This timescale is interestingly of the same order of the age of the disk stars. The mass function predicted by our mechanism is top-heavy. Additionally,
since TDEs are isotropic, our model predicts a random orientation for the disk of stars with respect to the plane of the galaxy and, due to the relatively high TDE rate, it can account for multiple disks of stars with uncorrelated orientations.

\end{abstract}

\keywords{Star formation -- Tidal disruption -  Supermassive Black Holes}

\section{Introduction}

The central region of the Milky Way has been the subject of considerable investigation for several decades. In addition to a supermassive black hole (SMBH) of about $4\times 10^6M_\odot$ \citep{Eckart1996,Boehle2016}, the central parsec region hosts over a hundred of young massive stars whose existence and formation channels are puzzling \citep{Levin2003,Genzel2003,Perets2009}, and whose organization in one \citep{Levin2003}, or possibly two, misaligned disks with respect 
to the galactic plane (e.g. \citealt{Paumard2006,Bartko2009,vF22}) is also of unclear nature.

Several models have been proposed
to explain the properties of these disk stars, from in-situ formation, to circular migration, to formation from a collapsing molecular cloud
(see Sec. 2.2). However, no model 
has been able to naturally explain all the main features. 

Here we propose a novel idea which can not only explain the  formation of the young stars in the close proximity of Sgr A*, but it also naturally predicts a clustering in a disk-like structure of random orientation with respect to the plane of the Galaxy, as well as the presence of multiple disks of different orientations with respect to one another.

Tidal disruption events (TDE) occur when a star wanders too close to the SMBH. The inferred observed TDE rates are of the order of $\sim 10^{-5}-10^{-4}\ \rm yr^{-1}$ per galaxy (e.g. \citealt{Gezari2008}), and is broadly consistent with theoretical estimates \citep[e.g.][]{Stone2016}. 

Among the several dozens of observed TDEs to date, only a handful of them  displays evidence for the presence of a jet \citep{vanvelzen2013,vanvelzen2016}, hence making the typical rate per galaxy of jetted TDEs to be on the order of $10^{-7}-10^{-6}$~yr$^{-1}$.

The feedback effect of jets  and, more generally, of conical outflows on their surrounding environment, and in particular on star formation, has been investigated by a number of authors \citep{Begelman1989,Silk2005,Gaibler2011,ga12,Nayakshin2012,Zubovas2014,Bieri2015,Bieri2016,zu19}. Of particular interest to this work, the numerical simulations by \citet{Gaibler2011} (in the context of AGN feedback) showed that compression in the plane perpendicular to the jet leads to enhanced star formation in a disk-like structure.

Here we propose that such mechanism may have operated in the vicinity of Sgr A*, where the jet is due to a TDE.
Multiple episodes of TDEs can lead to multiple disks in randomly oriented directions. We develop our idea in more detail as follows. Sec.~2 describes the observational properties of the two disks of stars and provides a summary of the main models proposed to explain them. The general properties of TDEs, both from theory and observations, are summarized in Sec.~3. We discuss the application of the jet-induced 
star formation mechanism to the Galactic center in Sec.~4. We summarize our ideas and findings in Sec.~5.

\section{Dynamics of the young stars in the Galactic center}

\subsection{Observations of the young stellar populations}

The dynamics of the young star cluster in the near vicinity of Sgr A* had been studied for over two decades. An investigation  by \citet{Genzel2000} revealed that most of the stars move clockwise.
A follow up analysis by \citet{Levin2003} of a sample of 13 stars within 0.4~pc from the Galactic center 
further revealed a puzzling feature: 10 of them lie within a thin disk inclined with respect to the galactic plane. The presence of a second disk with additional young stars and a different orientation was later suggested  \citep{Genzel2003}.

A detailed spectroscopic analysis by \citet{Paumard2006} confirmed the presence of the two disks, both misaligned with respect to the Galactic plane {and with an average aspect ratio between the scale height and the radius of $|h|/R \simeq 0.14$}.
The clockwise system, consistent with the one originally discovered by \citet{Levin2003}, was further populated with the identification of 36 stars at a distance of $\sim 0.04-0.08$~pc. Another counterclockwise system, less populated with 12 stars, was identified at larger distances, with an outer radius of about 0.5~pc. 

A proper motion analysis, based on 11 years of data from the Keck telescope,  was later performed by \citet{Lu2009}. For their primary sample comprising a cluster of 32 young stars within 0.14~pc, they confirmed the motion within a disk at a high inclination rotating in a clockwise sense. Their data did not however reveal a significant presence for a second disk.

More recently, a detailed analysis of the central star cluster was presented
by \citet{vF22}, who identified several kinematic features. While two of them were identified as the previously reported clockwise and counter-clockwise disks, two other prominent overdensities in angular momentum were 
reported. If further analysis of these features reveal new disk structures, it will be very interesting for the proposed formation mechanism. In the following subsection we will review previous models proposed to explain disk-like stars systems in the SMBH vicinity.

\subsection{Possible origins of the disk stellar populations}

The observations of the two disks of stars require two separate episodes of star formation, which are however allowed to be separated by $\simeq 1$~Myr from each other \citep{Paumard2006}. 
Several models have been suggested for the origin of these disk stars, and generally divided into two broad classes: in-situ formation, or
migration after formation in a far away cluster.

In-situ formation could be challenging due to the strong tidal forces, but tidal disruption of a molecular cloud could form a thin accretion disc of around $10^5 M_\odot$, which later fragments due to its own self gravity \citep{Levin2003,Genzel2003,Nayakshin2005, Paumard2006, Nayakshin2007, bonnell2008}. Hydrodynamical simulations of cloud-cloud collisions also reproduce some of the observed features \citep{hobbs2009}. 

An alternative possibility is the inspiraling cluster scenario \citep{Gerhard2001,McMillan2003,Portegies2003,Kim2003,Gurkan2005}, according to which stars formed in a massive cluster located far enough away to escape tidal disruption, and then they migrate inwards due to dynamical mechanisms. 
Regardless of the formation mechanism, a plethora of dynamical mechanisms has been invoked to explain the unique orbital configurations of the disk stars and, more generally, the central young star cluster (see e.g.  \citealp{alexander17} for a recent review and references).

\section{Tidal Disruption Events}

Tidal disruption events result from close approaches of stars to a BH. If a star of mass $M_*$ and radius $R_*$ approaches a BH of mass $M_{\rm BH}$, it will be tidally torn apart if it gets close enough to the BH that its self-gravity is overcome by the BH's tidal force.
The distance at which the two forces are comparable is the tidal radius, $R_t\approx (M_{\rm BH}/M_*)^{1/3} R_*$. After the disruption, about half of the stellar debris become unbound, while the other half returns to the BH and accretes at high rates. The accretion rate has a peak followed by a decline in time roughly as $\sim t^{-5/3}$ \citep{Rees1988}. 
Of special interest to the idea proposed in this work are the rates of these events, the typical energy released in each event, and the fact that a subset of TDEs displays evidence for jetted emission with inferred jet sizes comparable with the distance scale of the disks of  stars around Sgr A*.

The inferred TDE rates are of the order of $\sim 10^{-5}-10^{-4}\ \rm yr^{-1}$ per galaxy (e.g. \citealt{Donley2002,Gezari2008,vanVelzen2014}),
which matches theoretical modelling of nuclear clusters dynamics \citep{Magorrian1999,Wang2004, bar-or2016}. { The rates further display a dependence on the SMBH mass, being more prominent around lower mass SMBHs \citep{Stone2016,Broggi2022}}.

Fits to 14 well monitored TDE light curves by \citet{Mockler2021} yielded bolometric radiative energies varying between $\sim $~a few $\times 10^{50}-10^{53}$~ergs.  In most cases the radiative energy was found to fall short of the expectations for accretion of half a solar mass of material, 
perhaps due to the fact that a large fraction of the rest-mass energy is carried away by outflows \citep{Metzger2016}. 
{The amount of rest-mass energy which is converted into jet power depends on the properties of the accretion flow, and in particular on its accretion rate onto the SMBH. The calculations by \citet{Piran2015} for SMBHs of mass in the range $\sim 10^6-10^7 M_\odot$ and spin parameter $a\gtrsim 0.3$, find that the jet power is $\gtrsim 10^{45}$~erg~s$^{-1}$ up to timescales of several hundreds of days. Interestingly, note that recent observations with the Event Horizon Telescope suggest a spin $a >0.5$ for Sgr~A* \citep{EHT2022b}.  }

Given the typical TDE rates of $10^{-4}-10^{-5}\ \rm yr^{-1}$ and the few per cent of jetted TDEs \citep{vanvelzen2016}, we can conservatively estimate that $1\%$ of TDEs produce jets, which leads to jetted TDEs around Sgr A* of $\sim 0.1-1$ per million years. This is an especially interesting rate for our proposed model. 

The physical scale out to which a TDE jet can potentially trigger star formation is clearly connected to the length scale of the jet. 
Detailed radio observations of a few well-monitored events has allowed
such estimates to be made.
Under the assumption of energy equipartition (as for gamma-ray burst jets), \citet{Alexander2016} found that the jet had reached a distance of about $0.03$~pc  at the latest observation time of 381~days. The outflow from a candidate TDE reported by \citet{Somalwar2022} had an estimated radius of $\sim 0.7$~pc. 
A comparable or larger scale was estimated in the case of TDE AT2018hyz, where a delayed outflow was observed, and its analysis (under the assumption of jetted emission) led to derive a jet radial extent of parsec scale \citep{Cendes2022}. Especially interesting is the transient Arp 299-B AT1, where the jet is resolved in radio images, hence directly yielding a scale measurement, which, after a few years, reaches $\sim$~pc \citep{Mattila2018}.

\section{Connecting TDEs to star formation around Sgr A* }

\subsection{Jet-induced star formation}

Feedback on star formation has been shown to operate as a result of a variety of explosive events. 
Particularly important is the role of supernova shocks, which are shown to trigger and enhance star formation (e.g. \citealt{Chiaki2013}). Jets from AGNs and, more generally, galactic outflows are also believed to enhance star formation. In fact,  observational evidence of star formation is present in galactic outflows \citep{ma17}, and
 a link between jets and star formation is also suggested by a high  occurrence of young stars in compact radio sources \citep{Dicken2012}. 

Several theoretical and numerical studies in the context of galactic outflows and AGNs support the observations. Common findings are that jets and outflows trigger star formation on a very short timescale by overpressuring gas clouds
\citep{Begelman1989,Silk2005,Gaibler2011,ga12,Nayakshin2012,Zubovas2014}.
On the other hand,
feedback of TDEs on star formation has received much less attention, despite the fact that
the effect of recent star formation on the enhancements of TDE rates has been linked in post starburst galaxies \citep{hinkle21, bort22}. 
\cite{zu19} studied the properties of outflows driven by TDE-powered AGNs in dwarf galaxies, and found that these outflows may have significant influence on their host galaxies. Similarly to the other studies in galactic context,
this work found that
the outflowing gas, with its large pressure, can compress denser clumps of the ISM and enhance star formation 
 in the plane perpendicular to the jet.

 \subsection{Star formation from clouds overpressurized by a TDE jet cocoon}

 The formation of stars within a small region from the central SMBH requires densities high enough that the clumps are not tidally disrupted \citep{Phinney1989}. The Roche limit for a cloud of density $\rho$ at a distance $r$ from the SMBH is  (see e.g. \citealt{Sanders1998})

 \begin{equation}
 \rho >  1.5\times 10^{-13} \left(\frac{M_{\rm SMBH}}{10^6 M_\odot}\right)\left(\frac{0.1~{\rm pc}}{r}\
\right)^3\,{\rm g}~{\rm cm}^{-3}.
 \label{eq:rhoH}
 \end{equation}

 Here we follow the original suggestion by \cite{Begelman1989} that jets surrounded by cocoons expand sideways and shock-compress gas, triggering star formation from over-pressured clumps. Their original analysis is applied to the inter galactic medium (IGM) and to observations of Cygnus A, where both the bow shock head and the overpressurized cocoon are observed in radio. We will rescale the jet to Galactic sizes and interstellar  medium (ISM), which will result in much larger pressures, both for the cocoon and the ISM.

 \begin{figure}
     \centering
     \includegraphics[width=8.5cm]{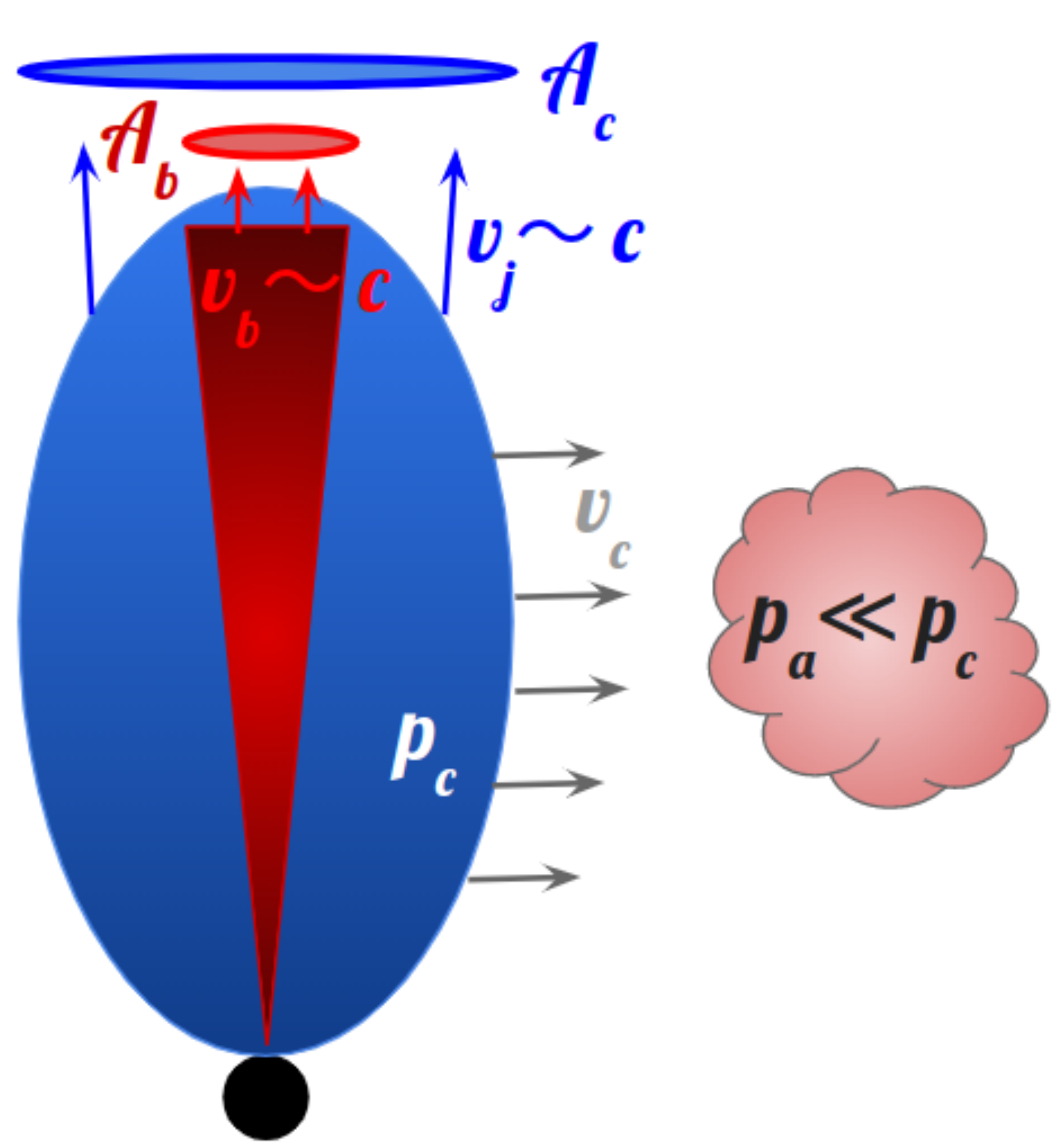}
     \caption{Sketch of the jet/cocoon system, as the over-pressured cocoon engulfs gas clouds.}
     \label{fig:1}
 \end{figure}
 Consider a relativistic jet of velocity $v_j \approx c$, surrounded by a cocoon. The edge of the cocoon carves into the ISM with velocity $v_b$. The observed hot-spots and bow shocks of radio sources { (i.e. \citealt{Carilli1988})} suggest that the hot-spot area under the bow shock, $A_b$, is substantially smaller than the  cocoon's cross sectional area, $A_c$. Figure \ref{fig:1} depicts a sketch of the jet geometry. The hot-spot velocity is derived by balancing the jet thrust $L_j/v_j$ and the ISM ram pressure force $\rho_a v_b^2 A_b$, where $L_j$ is the jet power and $\rho_a$ is the ISM density, yielding \citep{Begelman1989}
 \begin{equation}
     v_b = \left(\frac{L_j}{\rho_a v_j^3 A_b} \right)^{1/2} v_j\,. \label{eq:vb}
 \end{equation}
 \cite{Begelman1989} treat $v_b$ as a free parameter and take $A_b$ and $A_c$ from observations of Cygnus A.
 Here we will adopt a value of the ratio $w_c/l_c$ between the cocoon width and length in the range of { $\sim 1/2-1/6$} as 
estimated for Cygnus~A
  \citep{Begelman1989}, and a jet length on the order of a fraction of parsec as measured for a few TDE jets (see refs in Sec. 3). This allows us to estimate the cocoon cross sectional area $A_c$.

 The condition $v_b \lesssim v_j\sim c$ leads to a lower limit for the bow shock cross section:
 \begin{equation}
     A_{b}\gtrsim \frac{L_{j}}{\rho_{a}v_{j}^{3}}\equiv A_{b,{\rm lim}}=3.9\cdot10^{-4}L_{45}\,\rho_{-20}^{-1}\,\,{\rm pc^{2}}\,. \label{eq:Ab}
 \end{equation}
 Here the jet power is normalized to $L_{45}=L_j/10^{45}\ \rm erg\ s^{-1}$ and the ISM density  to $\rho_{-20} = \rho_a / 10^{-20}\ \rm g\ cm^{-3}$.

For a cocoon length $l_c=0.5~{\rm pc}$ and width $w_c\sim l_c/6\sim 0.083$~pc,
the cross sectional area of the orthogonal expansion is $A_c=\pi w_{c}^{2}= 0.02\ {\rm pc^{2}\gg}A_{b}.$
The cocoon expands transversally to the jet direction, with a velocity
\begin{equation}
v_c\sim \frac{\sqrt{A_c}}{t}\,,
\label{eq:vc}
\end{equation}
where the timescale evolution of the cocoon is given by \citep{Begelman1989}
\begin{equation}
    t\sim\left({\frac{\rho_a}{L_jv_j A_b}}\right)^{1/4}A_c\,. \label{eq:t}
\end{equation}
Finally, the pressure inside the cocoon can be estimated as
\begin{equation}
p_{c}\sim \rho_{a}v_{c}^{2}\sim
\rho_a\frac{A_c}{t^2}\sim \rho_a^{1/2}L_j^{1/2}v_j^{1/2}A_b^{1/2}A_c^{-1}.
\label{eq:pc}
\end{equation}
Numerically evaluating this expression with $v_j\sim c$,
$A_{c,0.02} = A_c/(0.02\ \rm pc^{2})$ and $A_b=A_{b,{\rm lim}}$, we obtain 
\begin{equation}
p_{c}\gtrsim \, 0.17\,\rho_{-20}^{1/2}\,L_{45}^{1/2}\,A_{c,0.02}^{-1}\,A_{b,{\rm lim}}^{1/2}\, {\rm \ dyn\ cm^{-2}}\,.\
 \label{eq:pc_num}
\end{equation}
The cocoon pressure needs to be compared with the pressure inside the cloud, which is in equilibrium with the ISM at \
 $p_a\sim \rho_a c_s^2$.  For a temperature $T=10^3\ \rm K$ and atomic  hydrogen composition, the sound speed is $c_s\
\sim 3$~km~s$^{-1}$, yielding $p_a \sim 10^{-9}$ dyn cm$^{-2}$\,. 
After the shock passes through the cloud, the pressure changes from $p_a$ to $p_c\gg p_a$.
The shocked clump can cool efficiently if the cooling timescale is shorter than the travel time of the shock through the clump. As a rough estimate of the conditions in our problem, we use the cooling function of post-shocked gas by \citet{Sgro1975}, yielding the ratio for the cooling time to shock travel time
\begin{align}
R = &4.6 \times 10^{-12}\alpha^{2.5} \left(\frac{n_{a}}{n_{cl}}\right)^{3.5} \nonumber \\ \times & \left(\frac{v_c}{{\rm km}~{\rm s}^{-1}}\right)^5 \left(\frac{n_a}{{\rm cm}^{-3}}\right)^{-1}
\left(\frac{d}{\rm pc}\right)^{-1}
\,,
\label{eq:ratio}
\end{align}
where $n_a$ and $n_{cl}$ are respectively the number densities of the ambient medium and of  the pre-shocked clump, 
$d$ is the clump size, 
$v_c$ is the shock velocity,
and $\alpha$ is a numerical factor dependent on the density ratio (varying between 1 and 4.4 as the density contrast $n_{cl}/n_a$ varies between 1 and 100).  As a representative example, let us consider a jet of luminosity $L_{45}=0.7$ expanding in an ambient medium of density $n_a=10^3$~{cm}$^{-3}$, and a clump  size of $\sim 0.04$~pc. Then we find $R\lesssim 1$ for a clump density\footnote{Overdensities of this order of magnitude have been observed in the close environment of Sgr A* also at the current time (i.e. \citealt{Yusef2017}). } $n_{cl} \gtrsim 5\times 10^{6}$~cm$^{-3}$ 
(corresponding to a clump of mass $\gtrsim 40 M_\odot$).  Under these conditions the shock is radiative\footnote{If the shock is not radiative, the same general ideas still apply, but the compression factor will be smaller and hence the star formation efficiency will be reduced.}, and the density in the cloud increases by the ratio $p_c/p_a \sim  10^8$, while its Jeans mass
 is reduced by a factor $(p_a/p_c)^{1/2}$. The  jump in density by $\sim 8$ orders of magnitude prevents the cloud from tidal disruption as it collapses to become a star.
Note that this mechanism tends to favour the formation of more massive stars, hence a top-heavy star population.  Additionally, as shocks also impart a kick to the shocked clumps (i.e. \citealt{Sgro1975}), the newly formed stars may acquire eccentricity in their orbits.

Last, we note that, while jets enhance star formation in the plane perpendicular to their
 direction, as discussed above, on the other hand they suppress it
 along them, further contributing to a disk-like star
 clustering. In fact, the strong soft X-ray/UV flux from the jet heats
 up and ionizes the ISM in conical regions centered along the jet
 axis. For a nominal UV energy of $E_{\rm TDE}=10^{50}E_{50}$~erg, a
 number of ionizing photons 
 $N_{\rm ph} \sim E_{\rm TDE} /{h\nu_{\rm ion}}\sim 4.6\times 10^{60}~E_{50}$ (using a photoionization energy of 13.6~eV)
will ionize a
 conical region up to a distance $D_{\rm ion}\,\approx\, 5.6\,
 (n_a/10^3{\rm cm}^{-3})^{-1/3}\,\tan\theta_{\rm jet,45}^{-2/3}\,E_{50}\, {\rm pc},$ where
  $\theta_{\rm jet,45}=\theta_{\rm jet}/45^\circ$ is a conservative value for the
 angular size of the ionizing source (since jets are likely to be smaller in angular size).  Heating of the gas raises the Jeans mass
 for star formation, while at the same time further preventing cooling
 of the clouds\footnote{This additional suppression will however last on the order of the cooling time $t_{\rm cool}\sim 100 (T/10^5{\rm K})n_3^{-1}$~yr (assuming optically thin gas).}.  Above a temperature of $\sim 10^4$~K, where hydrogen
 becomes ionized, the opacity becomes very high, hence effectively
 preventing cooling in the full cone. On the other hand, in the plane
 compressed by the jet cocoon, star formation would occur in the
 opacity gap $\sim 2000$K -- $10^4$~K where compressed gas is able to
 cool on a timescale shorter than the dynamical time (see discussion
 in \citealt{Zhu2009} and \citealt{nayakshin2018}).  These colder,
 shock compressed clumps would form the starting point for each 
 disk of stars.

Hence, to summarize, a TDE jet is expected to suppress star formation
in a conical region around its axis due to its strong
ionization/heating flux, while at the same time triggering it in the
perpendicular plane due to the high cocoon pressure which shock-compresses  ISM clouds above their Roche limit for tidal disruption by the SMBH. 
{ We however remark that a quantitative predicton of the thickness of
the disk of stars formed via this mechanism can only be made via
numerical simulations of the process, accounting both for the heating
radiation from the jet, as well as for the velocity distribution of
the laterally expanding cocoon.}

Since TDEs are
isotropic, the disk-like region of compressed material has a high
chance of being misaligned with the galactic disk. For the same
reason, multiple (jetted) TDEs would lead to enhanced star formation
in disks with different orientation from one another.
{Additionally, we note that for a nominal rate of jetted TDEs of one every few Myr as suggested by current observations (but obviously with the expectation of a stochastic distribution in time), the prediction of our model is that the two most prominent disks would be naturally associated with the last two jetted TDEs. Since our scenario favours a top-heavy mass function, disk stars from earlier TDEs triggering episodes are expected to quickly disappear due to the short lifetime of massive stars (for a star of $\sim 30 M_\odot$ the lifetime is about 2~Myr).  }

\section{Summary}

The two misaligned disks of young stars at sub parsec scales around Sgr A* are of unclear origin. Here we made a novel proposal for their formation, as star-forming bursts events triggered by 
TDEs by the SMBH Sgr A* itself. 
There are several attractive features of this idea which we have discussed, and which we summarize below:
\begin{itemize}

\item
Jet-induced star formation has been observed on a galactic scale, and this positive feedback has been confirmed by numerical simulations.
\item 
TDE rates are estimated to be on the order of $10^{-5}-10^{-4}$~yr$^{-1}$, and jetted TDEs are likely to be a percent fraction of them, hence making the jetted TDE rate $\sim 10^{-7}-10^{-6}$~yr$^{-1}$.
\item 
The young star population has an age estimated between 1-10~Myr, hence compatible with the rate of jetted TDEs.
\item
Observations of TDE outflows have revealed a radial extension up to pc scale, hence comparable to the size of the observed two disks of young stars.
\item
TDEs emit a large output of ionizing radiation; jetted TDEs will heat up and ionize the medium along cones of ISM, hence suppressing star formation in those regions.
\item
With analogy to jets studied in galactic outflows, TDEs jets are expected to enhance star formation in a plane perpendicular to the jet, due to the enhanced pressure of the jet on the ISM. The cocoon surrounding a TDE jet can be over-pressured compared to the ISM pressure by $\sim 7-8$ orders of magnitudes. Clumps can be extremely shock-compressed by the jet cocoon,  and hence resist tidal disruption in the close Sgr A* vicinity, thus collapsing to form stars. 
A star population produced via our proposed mechanism is predicted to be top-heavy.
\item
TDEs are randomly oriented with respect to the plane of the galaxy; hence the disk (or disks) of TDE-induced stars is most likely going to be misaligned with respect to the plane of the galaxy. 

\end{itemize}

The idea proposed here, while compelling for the reasons described above, will need to be tested and quantified via detailed numerical simulations, and we hope that our work will serve as a motivation in this direction. While we have focused our study to the special case of our Galactic center,
TDE feedback on the ISM is a problem that concerns pretty much every galaxy. In fact, while
in a typical galaxy the rate of TDEs is smaller than that of other explosive phenomena like SNe, the fact that TDEs occur always at the same location in the very center of a galaxy\footnote{TDEs can also occur due to random encounters between stars and stellar-mass black holes, however, these will be at varying locations.} around its SMBH, makes TDE feedback potentially important in self-regulating star formation and hence the TDE rate itself. Since 
TDEs are more prominent around lower mass SMBHs,  our TDE-feedback on star formation in the innermost regions  can
potentially be relevant also to Intermediate Mass BHs  in Globular Clusters.

\begin{acknowledgements}

 RP thanks Scott Tremaine for early encouragement to pursue this idea and Yuri Levin for a very
 informative discussion. RP also
 kindly acknowledges support by NSF award AST-2006839. EG thanks Ryosuke Hirai for useful discussions. We further thank Yuri Levin { and an anonymous referee} for very helpful comments on the manuscript.

\end{acknowledgements}

\bibliography{references}
\bibliographystyle{aasjournal}


\end{document}